\documentclass[onecollarge]{svjour2}       
\smartqed  
\usepackage{psfrag,graphicx}
\journalname{Journal of Statistical Physics}
\begin{document}

\title{document}
\title{Perturbation theory for a stochastic process with Ornstein-Uhlenbeck noise}

\author{Michael Wilkinson}

\institute{Michael Wilkinson \at
              Department of Mathematics and Statistics, \\
              The Open University,
              Walton Hall, \\
              Milton Keynes, MK7 6AA, \\
              England.\\
              \email{m.wilkinson@open.ac.uk}      \\
}

\date{Received: date / Accepted: date}

\maketitle

\begin{abstract}

The Ornstein-Uhlenbeck process may be used to generate a noise signal with a finite correlation
time. If a one-dimensional stochastic process is driven by such a noise source, it may be analysed by solving
a Fokker-Planck equation in two dimensions. In the case of motion in the vicinity of an attractive fixed
point, it is shown how the solution of this equation can be developed as a power series. The coefficients
are determined exactly by using algebraic properties of a system of annihilation and creation operators.


\end{abstract}

\maketitle

\section {Introduction}
\label{sec: 1}

Many physical processes are modelled by adding noise to a dynamical
system. This paper discusses a one-dimensional example where noise is added to a system with an
attracting fixed point. With a suitable change of variables the fixed point is positioned at the origin
and the system may be expressed in the form
\begin{equation}
\label{eq: 1.1}
\dot x=-x-\epsilon g(x) + f(t)
\end{equation}
where $f(t)$ is a random noise with statistics
\begin{equation}
\label{eq: 1.2}
\langle f(t)\rangle=0\ ,\ \ \ \langle f(t_1)f(t_2)\rangle=C(t_1-t_2)
\end{equation}
(throughout this paper $\langle X\rangle$ denotes the expectation value of any random variable $X$).
In (\ref{eq: 1.1}) the non-linearity of the stochastic process is represented by the function $g(x)$, which is
assumed to satisfy $g(0)=g'(0)=0$, with a multiplier $\epsilon$ which will be used as a perturbation parameter.
In the case where $f(t)$ is a white noise signal, with correlation function
$C(\Delta t)=2D\delta (\Delta t)$, the probability density for the solution of
(\ref{eq: 1.1}) satisfies a Fokker-Planck equation:
\begin{equation}
\label{eq: 1.3}
\frac{\partial P}{\partial t}=\frac{\partial }{\partial x}[(x+\epsilon g(x))P]+D\frac{\partial ^2 P}{\partial x^2}
\ .
\end{equation}
In one dimension it is easy to obtain exact steady state solutions of this equation. When $\epsilon=0$, it
is also possible to determine the propagator and use this to determine correlation functions exactly (this case
is known as the Ornstein-Uhlenbeck process \cite{Uhl+30}, described in \cite{vKa81,Ris89}). The exact results which are available when $\epsilon=0$ can be used as a basis for a perturbation expansion in $\epsilon$ (an example of the application of this approach is described in \cite{Meh+04}). When the
noise is not delta-correlated however, it is much more difficult to analyse equation (\ref{eq: 1.1}). This paper
considers a particular case where the noise in (\ref{eq: 1.1}) is not delta-correlated, but where it
is nevertheless possible to analyse the statistics of the solutions by means of a Fokker-Planck equation.
This is possible if the noise is itself generated by a dynamical process which is driven by a white
noise signal.

The following model will be analysed:
\begin{eqnarray}
\label{eq: 1.4}
\dot x&=&-x-\epsilon g(x)+y
\nonumber \\
\dot y&=&-\omega y +\omega \xi(t)
\end{eqnarray}
where $\xi(t)$ is a white-noise signal with statistics
\begin{equation}
\label{eq: 1.5}
\langle \xi(t)\rangle=0 \ ,\ \ \ \langle \xi(t_1)\xi(t_2)\rangle=2\delta(t_1-t_2)
\ .
\end{equation}
The variable $y(t)$ is not influenced by $x(t)$. It is a coloured noise signal which is generated by an Ornstein-Uhlenbeck process, which has an exponential correlation function \cite{Uhl+30,vKa81,Ris89}
\begin{equation}
\label{eq: 1.6}
\langle y(t_1)y(t_2)\rangle =\omega\exp(-\omega \vert t_1-t_2\vert)\ .
\end{equation}
This approach to modelling systems with coloured noise generated by an Ornstein-Uhlenbeck process was previously
considered by Fox {\em et al} \cite{Fox+88}, who pointed out that the numerical simulation of systems with coloured noise is facilitated by using an Ornstein-Uhlenbeck process to generate the noise, and by Risken \cite{Ris89}, who noted that it allows process with coloured noise to be modelled using a Markovian process. The process represented by equations (\ref{eq: 1.4}), (\ref{eq: 1.5}) and (\ref{eq: 1.6}) approaches the process described by (\ref{eq: 1.3}) in the limit as $\omega \to \infty$, with the diffusion constant $D=1$ (a scaling of $x$ and $t$ can always be applied so as to set $D=1$).

In many applications the precise form of the correlation function of the noise is not known, and the analysis of this model is sufficient to understand the effect of the noise having a finite correlation time. In cases where the precise form of the correlation function of the noise is significant, the approach which is developed here can be extended to more general correlation functions by generalising the dynamical process which is used to smooth the white-noise signal.

The joint probability density for $x$ and $y$ satisfies the Fokker-Planck equation
\begin{equation}
\label{eq: 1.7}
\frac{\partial P}{\partial t}=\frac{\partial}{\partial x}[(x-y+\epsilon g(x))P]+\omega\frac{\partial }{\partial y}(yP)+\omega^2\frac{\partial^2P}{\partial y^2}
\end{equation}
which will be written as
\begin{equation}
\label{eq: 1.8}
\frac{\partial P}{\partial t}=[\hat {\cal F}_0+\epsilon \hat {\cal F}_1]P
\end{equation}
(a \lq hat' over a symbol will be used to denote a differential operator).
Our objective is to obtain the solution of the steady-state Fokker-Planck equation in the form of a
power series
\begin{equation}
\label{eq: 1.9}
P(x,y)=\sum_{j=0}^\infty \epsilon^j P_j(x,y)
\ .
\end{equation}
Inserting this into (\ref{eq: 1.8}) gives the recursion relation $P_{j+1}(x,y)=-\hat {\cal F}_0^{-1}\hat {\cal F}_1 P_j(x,y)$. It is not immediately clear how the inverse $\hat {\cal F}_0^{-1}$ can be computed or whether this equation gives a meaningful expression for $P_{j+1}$. The operator $\hat {\cal F}_0^{-1}$ will be defined by analysing the spectrum of $\hat {\cal F}_0$. It is clear that there is an eigenfunction of $\hat {\cal F}_0$ with eigenvalue equal to zero, which is the steady-state solution of (\ref{eq: 1.7}) when $\epsilon=0$. This suggests that the inverse $\hat {\cal F}_0^{-1}$ is ill-defined. However, it will be shown that the terms in the series are in fact well-defined. Moreover, the coefficients can be determined exactly by using the algebraic properties of a system of raising and lowering operators, which are defined in section \ref{sec: 2}, before developing the perturbation theory in section \ref{sec: 3}. The use of annihilation and creation operators to treat Fokker-Planck equations is discussed in the book by Risken \cite{Ris89}, but the usual approach is not applicable to the problem which is treated here, and a different method is required. This point is considered in section \ref{sec: 4}.

A motivation for developing this approach was to study the Lyapunov exponents for inertial particles suspended in random fluid flows. This problem can be transformed into determining the expectation value of a stochastic variable such as (\ref{eq: 1.1}), where $f(t)$ represents the velocity gradient of the fluid at the position occupied by the particle \cite{Wil+03,Meh+04}. The technique which is developed here will be applied to the calculation of the Lyapunov exponent in a companion paper \cite{Wil09}. In this work, however, the method will be developed in a general setting, which will surely find applications in other areas.

\section {Unperturbed eigenfunctions and spectrum}
\label{sec: 2}

The steady state solution of the Fokker-Planck equation (\ref{eq: 1.7}) for the $\epsilon=0$ problem is a
Gaussian function
\begin{equation}
\label{eq: 2.1}
P_0(x,y)=\exp\left[-\frac{1}{2\omega^2}(1+\omega)\left((1+\omega)x^2+y^2-2xy\right)\right]
\ .
\end{equation}
Writing $P=fP_0$, where $P$ is an eigenfunction of the Fokker-Planck equation
satisfying $\hat {\cal F}_0 P=\lambda P$, it is found that $f$ satisfies
\begin{equation}
\label{eq: 2.2}
(x-y)\partial_x f+[2(1+\omega)x-(2+\omega)y]\partial_yf+\omega^2\partial^2_yf=\hat F_0 f=\lambda f
\ .
\end{equation}
The operator $\hat F_0$ defined in (\ref{eq: 2.2}) maps polynomials to polynomials, and the generalised eigenfunctions of $\hat F_0$ satisfying $\hat F_0 f(x,y)=\lambda f(x,y)$ must be polynomials in $x$ and $y$. Clearly $f_0(x,y)=1$ is an eigenfunction with $\lambda=0$ and by inspection the linear functions $f_1(x,y)=(1+\omega)x-y$ and $f_\omega(x,y)=2x-y$ have eigenvalues $\lambda=-1$ and $\lambda=-\omega$ respectively. This observation motivates the definition of \lq lowering operators', $\hat B_1$ and $\hat B_\omega$, which satisfy $\hat B_1 f_\omega=0$ and $\hat B_\omega f_1=0$. Such
operators can be constructed as linear combinations of derivatives, $\partial_x=\partial/\partial x$ and $\partial_y=\partial/\partial y$. By inspection, it is possible to construct a set of raising and lowering operators
which satisfy the commutation relations:
\begin{equation}
\label{eq: 2.3}
\begin{array}{lll}\left[\hat F_0,\hat A_1\right]=-\hat A_1
&\quad&
\left[\hat F_0,\hat A_\omega\right]=-\omega \hat A_\omega
\cr
&&\cr
\left[\hat F_0,\hat B_1\right]=\hat B_1
&\quad&
\left[\hat F_0,\hat B_\omega\right]=\omega \hat B_\omega
\end{array}
\end{equation}
(where $[\hat A,\hat B]\equiv\hat A \hat B-\hat B\hat A$).
These relations imply that the eigenvalues are $\lambda_{n,m}=-(n+\omega m)$.
The required operators are
\begin{equation}
\label{eq: 2.4}
\begin{array}{lll}
\hat A_\omega=2x-y-\frac{\omega}{\omega+1}\partial_x+\frac{\omega(\omega-1)}{\omega+1}\partial_y
&\quad &
\hat A_1=(1+\omega)x-y-\frac{\omega^2}{\omega+1}\partial_x
\cr
&\quad &\cr
\hat B_\omega=\partial_x+(1+\omega)\partial_y
&\quad &
\hat B_1=\partial_x+2\partial_y
\ .
\end{array}
\end{equation}
The operators $\hat A_1$ and $\hat A_\omega$ are termed raising operators because they increase the quantum number $n$ and $m$ respectively. The operators $\hat B_1$ and $\hat B_\omega$ are termed lowering operators.

Rather than working with the polynomials which are eigenfunctions of $\hat F_0$ it is more convenient to work with eigenfunctions of the operator $\hat {\cal F}_0$ defined by (\ref{eq: 1.7}), (\ref{eq: 1.8}), satisfying $\hat {\cal F}_0 \phi_{nm}=-(n+\omega m)\phi_{nm}$.
The operators in (\ref{eq: 2.4}) can be transformed into raising and lowering operators for generating
eigenfunctions of $\hat {\cal F}_0$ directly. These are
\begin{equation}
\label{eq: 2.5}
\begin{array}{lll}
\hat \alpha_1=-\partial_x
&\quad &
\hat \alpha_\omega=-\partial_x+(\omega-1)\partial_y
\cr
&\quad &\cr
\hat \beta_1=\frac{\omega^2}{\omega^2-1}\left[\partial_x+2\partial_y\right]+x+\frac{1}{\omega-1}y
&\quad &
\hat \beta_\omega=\frac{\omega}{1-\omega^2}\left[\partial_x+(1+\omega)\partial_y\right]-\frac{1}{\omega-1}y
\  .
\end{array}
\end{equation}
These satisfy the commutation relations
\begin{equation}
\label{eq: 2.6}
\begin{array}{lll}
\left[\hat \alpha_1,\hat \beta_1\right]=-1
&\quad &
\left[\hat \alpha_\omega,\hat \beta_\omega\right]=-1
\cr
&\quad&\cr
\left[\hat \alpha_1,\hat \alpha_\omega\right]=0
&\quad&
\left[\hat \beta_1,\hat \beta_\omega\right]=0
\cr
&\quad&\cr
\left[\hat \alpha_1,\hat \beta_\omega\right]=0
&\quad&
\left[\hat \alpha_\omega,\hat \beta_1\right]=0
\end{array}
\end{equation}
and
\begin{equation}
\label{eq: 2.7}
\begin{array}{lll}
\left[\hat {\cal F}_0,\hat \alpha_1\right]=-\hat \alpha_1
&\quad&\
\left[\hat {\cal F}_0,\hat \beta_1\right]=\hat \beta_1
\cr
&\quad&\cr
\left[\hat {\cal F}_0,\hat \alpha_\omega\right]=-\omega \hat \alpha_\omega
&\quad&
\left[\hat {\cal F}_0,\hat \beta_\omega\right]=\omega \hat \beta_\omega
\ .
\end{array}
\end{equation}
The inverse relations to (\ref{eq: 2.5}) are:
\begin{equation}
\label{eq: 2.8}
\begin{array}{lll}
\partial_x=-\hat \alpha_1
&\quad&
\partial_y=\frac{1}{\omega-1}\left[\hat \alpha_\omega-\hat \alpha_1\right]
\cr
&\quad&\cr
x=\hat \beta_1+\hat \beta_\omega+\frac{\omega}{\omega^2-1}\left[\omega\hat\alpha_1-\hat \alpha_\omega\right]
&\quad&
y=(1-\omega)\hat \beta_\omega+\frac{\omega}{\omega^2-1}\left[2\omega \hat \alpha_1-(\omega+1)\hat \alpha_\omega\right]
\ .
\end{array}
\end{equation}
Expressing $\hat{\cal F}_0$ in terms of the raising and lowering operators yields
\begin{equation}
\label{eq: 2.9}
\hat {\cal F}_0=-\left(\omega \hat \alpha_\omega \hat \beta_\omega +\hat \alpha_1 \hat \beta_1\right)
\ .
\end{equation}
The notation $|\psi)$ will be used as a shorthand for a function $\psi(x,y)$ (this is an adaptation of the Dirac notation of quantum mechanics). The eigenfunctions $\phi_{n,m}(x,y)$ of
$\hat {\cal F}_0$ are denoted by vectors $|\phi_{n,m})$ and their eigenvalues are given by
\begin{equation}
\label{eq: 2.10}
\hat {\cal F}_0 \, |\phi_{n,m})=-(n+\omega m)\,|\phi_{n,m})
\ .
\end{equation}
Because $\hat {\cal F}_0$ is not self-adjoint, these eigenfunctions are not orthogonal. Neither
are they assumed to be normalised. The commutation relations (\ref{eq: 2.6}), (\ref{eq: 2.7}) for the raising and lowering operators
are consistent with the following relations defining how the eigenfunctions may be generated
by successive application of the raising operators:
\begin{equation}
\label{eq: 2.11}
\begin{array}{lll}
\hat \alpha_1 |\phi_{n,m})=|\phi_{n+1,m})
&\quad&
\hat \alpha_\omega |\phi_{n,m})=|\phi_{n,m+1})
\cr
\hat \beta_1 |\phi_{n,m})=n\,|\phi_{n-1,m})
&\quad&
\hat \beta_\omega |\phi_{n,m})=m\,|\phi_{n,m-1})
\ .
\end{array}
\end{equation}
The first two relations enable all other eigenstates to be generated by successive application of
the raising operators starting from the steady state $|\phi_{0,0})$. It will be assumed that $|\phi_{0,0})$ is normalised as a probability density, but the other eigenstates need not be normalised.

\section {Iteration of perturbation series expansion}
\label{sec: 3}

Consider the iteration of the perturbation series expansion (\ref{eq: 1.9}) starting from $|P_0)=|\phi_{0,0})$. Assuming that each term is expanded in terms of the eigenfunctions:
\begin{equation}
\label{eq: 3.1}
|P_j)=\sum_{n=0}^\infty \sum_{m=0}^\infty c^{(j)}_{n,m}\, |\phi_{n,m})
\ .
\end{equation}
The following considers how the perturbation series may be evaluated for the case where $g(x)$ is expressed as
a polynomial in $x$. The case $g(x)=x^2$ is considered explicitly, but more general polynomials are treated in exactly
the same way.

Successive terms in the series expansion are given by
\begin{equation}
\label{eq: 3.2}
|P_{j+1})=-\hat {\cal F}_0^{-1}\hat {\cal F}_1 \,|P_j)=\hat {\cal F}_0^{-1}\hat \alpha_1 \hat x^2 |P_j)
\ .
\end{equation}
The definition of the inverse $\hat {\cal F}_0^{-1}$ appears problematic, because one of the eigenvalues of $\hat {\cal F}_0$ is equal to zero. Note however that the operator $\hat {\cal F}_0^{-1}$ acts on a state which is multiplied by the raising operator $\hat \alpha_1$, so there is no division by zero. More explicitly, if a function $Q(x,y)$ is expressed in the form
\begin{equation}
\label{eq: 3.3}
Q(x,y)=\sum_{n=0}^\infty \sum_{m=0}^\infty q_{n,m}\,\phi_{n,m}(x,y)
\end{equation}
then, using (\ref{eq: 2.11}),
\begin{equation}
\label{eq: 3.4}
\hat {\cal F}_0^{-1}\hat \alpha_1 |Q)=-\sum_{n=0}^\infty \sum_{m=0}^\infty q_{n,m} \frac{1}{n+1+\omega m}|\phi_{n+1,m})
\end{equation}
which is well-defined. The operator $\hat x^2$ may be expressed in terms of raising and lowering operators via (\ref{eq: 2.8}), so that $\hat x^2|P_j)$ can be expressed as a linear combination of the form (\ref{eq: 3.3}) with known coefficients. It is, therefore, possible to iterate to determine the coefficients $c^{(j)}_{nm}$ in (\ref{eq: 3.1}), starting from $c^{(0)}_{nm}=\delta_{n0}\delta_{m0}$. The successive corrections $|P_j)$ are generated from (\ref{eq: 3.2}), using (\ref{eq: 2.8}), (\ref{eq: 2.11}) and (\ref{eq: 3.4}).

It is often moments of the distribution which are required, rather than the probability density itself. These can be obtained from the expansion (\ref{eq: 3.1}) if the moments of the eigenfunctions $|\phi_{n,m})$ are known. First, note that the raising operators are constructed from partial derivatives. It follows (using integration by parts) that, except for the case $n=m=0$,
\begin{equation}
\label{eq: 3.5}
\int_{-\infty}^\infty {\rm d}x \int_{-\infty}^\infty{\rm d}y\ \phi_{n,m}(x,y)=\delta_{n,0}\delta_{m}
\ .
\end{equation}
To determine other moments of the functions $|\phi_{n,m})$, such as $\langle x^N\rangle$, express $x$ and in terms of raising and lowering operators using (\ref{eq: 2.8}), so that $x^N\phi_{n,m}(x,y)$ may be expressed as a linear combination of eigenstates:
\begin{equation}
\label{eq: 3.6}
\hat x^N |\phi_{n,m})=\sum_{n'=0}^\infty \sum_{m'=0}^\infty W(N,n,m,n',m')|\phi_{n',m'})
\end{equation}
where the coefficients $W(N,n,m,n',m')$ are readily obtained from the
commutation relations of the raising and lowering operators. Upon integration over $x$ and $y$ the
only contribution comes from the term where $n'=m'=0$, so that
\begin{equation}
\label{eq: 3.7}
\langle x^N\rangle=\int_{-\infty}^\infty {\rm d}x \int_{-\infty}^\infty {\rm d}y\ x^N \phi_{n,m}(x,y)=W(N,n,m,0,0)
\ .
\end{equation}
As an example, consider the evaluation of $\langle x\rangle$:
\begin{equation}
\label{eq: 3.8}
\hat x\,|\phi_{nm})=n|\phi_{n-1,m})+m|\phi_{n,m-1})+\frac{\omega^2}{\omega^2-1}|\phi_{n+1,m})
+\frac{\omega}{1-\omega^2}|\phi_{n,m+1})
\end{equation}
so that the only cases where $W(1,n,m,0,0)\ne 0$ are $W(1,1,0,0,0)=1$ and $W(1,0,1,0,0)=1$. This means that if the probability density $P(x,y)$ is expressed in the form (\ref{eq: 3.3}) with coefficients $p_{n,m}$ then $\langle x \rangle= p_{1,0}+p_{0,1}$. From (\ref{eq: 3.4}) it can be seen that the application of $\hat {\cal F}_0^{-1}\hat \alpha_1$ to $|Q)$ yields a state for which the coefficient $q_{0,1}$ is equal to zero. It follows that the series expansion of $\langle x\rangle$ is
\begin{equation}
\label{eq: 3.9}
\langle x\rangle=\sum_{j=0}^\infty \epsilon^j\, c^{(j)}_{1,0}
\ .
\end{equation}
The coefficients can be determined using an algebraic manipulation package. They are equal to zero for even values of $j$. The first three non-zero coefficients are
\begin{eqnarray}
\label{eq: 3.10}
c^{(1)}_{1,0}&=&\frac{\omega}{\omega+1}
\nonumber \\
c^{(3)}_{1,0}&=&\frac {\omega^2\,(10\,\omega^2 + 27\,\omega + 15)}
{(\omega + 1)^3\,(1 + 2\,\omega)}
\nonumber \\
c^{(5)}_{1,0}&=&
\frac {10\,(72\,\omega^6 + 540\,\omega^5 + 1630\,\omega^4 + 2493\,\omega^3 + 2010\,\omega^2 + 807\,\omega + 126)\,\omega^3}
{(1 + 2\,\omega)^2\,(\omega + 1)^5\,(1 + 3\,\omega)\,(2 + \omega)}
\ .
\end{eqnarray}
The complexity of the coefficients increases rapidly as $j$ increases. In the limit as $\omega\to \infty$,  where the coefficients approach those for the case of a white noise signal, the coefficients are already known. As $\omega\to \infty$ the non-vanishing coefficients approach $1,5,60,1105,\ldots$, which is in agreement with setting $\Gamma=0$ in the results contained in \cite{Meh+04}.

\section {Concluding remark}
\label{sec: 4}

The technique described here may find applications in a wide variety of systems which can be modelled by stochastic differential equations. A companion paper will describe the application of
the method to the calculation of Lyapunov exponents for particles moving in a random fluid flow.

One technical point shown be remarked upon. There is a standard approach to linear Fokker-Planck equations in which the Fokker-Planck operator $\hat {\cal F}_0$ is transformed to a Hermitian operator ${\cal H}_0$, writing
\begin{equation}
\label{eq: 4.1}
\hat {\cal H}_0=\hat {\cal T}^{-1}\hat {\cal F}_0\hat {\cal T}
\end{equation}
where the transformation $\hat {\cal T}$ is a \lq gauge transformation', $\exp[\Phi(x,y)]$.
By a suitable choice of a phase function $\Phi(x,y)$ which is a quadratic form in $x$ and $y$ the non-Hermitian contributions to $\hat {\cal H}_0$ can usually be eliminated. This approach is discussed in \cite{Ris89}. The Hermitian form of the operator is convenient for subsequent calculations, because its eigenfunctions are orthogonal. Moreover, because $\hat {\cal H}_0$ is a quadratic form in position and momentum operators $x_i$ and $\hat p_i=-{\rm i}\partial_{x_i}$ respectively, $\hat {\cal H}_0$ can be transformed into a harmonic oscillator Hamiltonian, which allows standard harmonic oscillator raising and lowering operators to be used directly (this approach was used in \cite{Meh+04}). This method is, however, not applicable to the operator $\hat {\cal F}_0$ defined by (\ref{eq: 1.7}) and (\ref{eq: 1.8}), because the non-Hermitian components of $\hat {\cal H}_0$ cannot be eliminated by any choice of the coefficients of the quadratic form $\Phi(x,y)$.

There are two possible ways to avoid this difficulty. One is to explore whether the required results can be obtained using the set of non-orthogonal eigenfunctions of $\hat {\cal F}_0$ generated by the operators $\hat\alpha_1$ and $\hat\alpha_\omega$. This is the approach which has been adopted here. It has been shown that series expansions of $\langle x^N\rangle$ may be obtained using an expansion in terms of eigenfunctions which are neither orthogonal nor normalised.

In some other contexts it may still be desirable to transform $\hat {\cal F}_0$ to a Hermitian operator, so that orthonormal bases can be used. It is possible to make a more general transformation of the form (\ref{eq: 4.1}), which is closely related to the definition of the Weyl representation of metaplectic operators in quantum mechanics, as discusses in \cite{Meh+01}. To this end it is useful to define an operator
\begin{equation}
\label{eq: 4.2}
\hat {\cal M}_x(\alpha)=\int_{-\infty}^\infty {\rm d}X\ \exp(-X^2/2\alpha)\exp(-X\partial_x)
\end{equation}
and a similar operator $\hat {\cal M}_y(\alpha)$, in which $\partial_x$ is replaced by $\partial _y$. To understand the significance of this operator, consider its action upon $xf(x)$. In the following manipulations $\exp(-X\partial_x)$ is identified with its Taylor series and hence interpret it as a translation operator which shifts a function by $X$: $\exp(-X\partial_x)f(x)=f(x-X)$. Assuming that $f(x)$ is a normalisable and sufficiently smooth function,
\begin{eqnarray}
\label{eq: 4.3}
\hat {\cal M}_x x f(x)&=&\int_{-\infty}^\infty {\rm d}X\ \exp(-X^2/2\alpha)(x-X)f(x-X)
\nonumber \\
&=&\int_{-\infty}^\infty {\rm d}X\ \exp(-X^2/2\alpha)[x\exp(-X\partial_x)f-X\exp(-X\partial_x)f]
\nonumber \\
&=&x\hat {\cal M}_x f+\alpha \int_{-\infty}^\infty {\rm d}X\ \frac{\rm d}{{\rm d}X}\left[\exp(-X^2/2\alpha)\right]\exp(-X\partial_x)f
\nonumber \\
&=&x \hat {\cal M}_x f-\alpha \int_{-\infty}^\infty {\rm d}X\ \exp(-X^2/2\alpha)\frac{\rm d}{{\rm d}X}\exp(-X\partial_x)f
\nonumber \\
&=&x\hat {\cal M}_x f+\alpha \partial_x \hat {\cal M}_x f
\ .
\end{eqnarray}
Because $\hat {\cal M}_x(\alpha)$ commutes with $\partial_x$, (\ref{eq: 4.3}) implies two equivalent rules for commuting $x$ and $\hat {\cal M}_x(\alpha)$:
\begin{equation}
\label{eq: 4.4}
\hat {\cal M}_x(\alpha) x=(x +\alpha \partial_x )\hat {\cal M}_x(\alpha)
\ ,\ \ \
x \hat {\cal M}_x(\alpha)=\hat {\cal M}_x(\alpha)(x-\alpha \partial_x)
\ .
\end{equation}
Note that $\hat {\cal M}_x(\alpha)$ also commutes with $\partial_y$, and $y$.
The action of operators $\hat {\cal M}_x(\alpha)$ and $\hat{\cal M}_y(\beta)$ enables $\hat {\cal F}_0$ to be converted into a partial differential operator which is of second order in both $x$ and $y$. It is found that the application of the operator $\hat {\cal M}_x(\alpha)$ was not required to convert $\hat {\cal F}_0$ to self-adjoint form, however it is necessary to introduce a multiplication by a scalar function. A transformation operator of the form
\begin{equation}
\label{eq: 4.5}
\hat {\cal T}=\hat {\cal M}_y(\beta)\exp(-\Phi)
\end{equation}
can be used in (\ref{eq: 4.1}), where $\Phi(x,y)$ is a quadratic form $\Phi(x,y)=\frac{1}{2}(Ax^2+By^2+2Cxy)$. The parameters $A$, $B$, $C$, and $\beta$ can be chosen so that $\hat {\cal H}_0$ is self-adjoint.

{\sl Acknowledgement}. I am grateful to Michael Morgan for a careful reading of the manuscript.


\begin{thebibliography}{0}

%

\bibitem{Uhl+30}
G. E. Uhlenbeck and L. S. Ornstein,
{\sl On the theory of the Brownian motion},
{\it Phys. Rev.}, {\bf 36}, 823-41, (1930).
%
\bibitem{vKa81}
N. G. van Kampen,
{\em Stochastic processes in physics and chemistry}, 2nd ed.,
North-Holland, Amsterdam, (1981).
%
\bibitem{Ris89}
H. Risken,
{\em The Fokker-Planck equation: methods of solution and applications}, 2nd ed.,
Springer, Berlin, (1989).
%
\bibitem{Meh+04}
B. Mehlig and M. Wilkinson,
{\sl Coagulation by random velocity fields as a Kramers problem},
{\it Phys. Rev. Lett.}, {\bf 92}, 250602, (2004).
%
\bibitem{Fox+88}
R. F. Fox, I. R. Gatland, R. Roy, G. Vemuri,
{\sl Fast, accurate algorithm for numerical simulation of exponentially correlated coloured noise},
{\it Phys. Rev. A}, {\bf 38}, 5938-40, (1988).
%
\bibitem{Wil+03}
M. Wilkinson and B. Mehlig,
{\sl The path-coalescence transition and its applications},
{\it Phys. Rev. E}, {\bf 68}, 040101(R), (2003).
%
\bibitem{Wil09}
M. Wilkinson,
{\sl Lyapunov exponent for small particles in smooth one-dimensional flows},
submitted to {\it J. Stat Phys.}, (2009).
%
\bibitem{Meh+01}
B. Mehlig and M. Wilkinson,
{\sl Semiclassical trace formulae using coherent states},
{\it Ann. Phys. Leipzig}, {\bf 10}, 541-59, (2001).

\end{thebibliography}
\end{document}